\documentclass[preprint,3p, onecolumn,times]{elsarticle}

\usepackage{lineno}

\usepackage{amsmath,amssymb,amsfonts}
\usepackage{algorithmic}
\usepackage{graphicx}
\usepackage{textcomp}
\usepackage{xcolor}
\usepackage{listings}
\usepackage{hyperref}
\usepackage{tabularx}
\usepackage{geometry}
\usepackage{multirow}
\usepackage{array}
\usepackage{balance}
\usepackage{booktabs}
\usepackage{url}
\usepackage{tcolorbox}
\usepackage{comment}
\lstdefinestyle{solidity}{
  language=C,
  basicstyle=\ttfamily\scriptsize,
  keywordstyle=\color{blue}\bfseries,
  commentstyle=\color{green!60!black},
  stringstyle=\color{red},
  numbers=left,
  numberstyle=\tiny\color{gray},
  numbersep=5pt,
  frame=none,                  
  xleftmargin=0.5cm,
  xrightmargin=0cm,
  aboveskip=0pt,
  belowskip=0pt,
  breaklines=true, 
  showspaces=false,
  showstringspaces=false,
  tabsize=2,
  escapeinside={\%*}{*)},
  morekeywords={function,returns,uint256,address,override,public,external,virtual,require,unchecked,return}
}
\usepackage{minted}

\usepackage{listings}
\usepackage{xcolor}
\usepackage[byauthor, separator = {;}]{credits}
\credit{Ermanno Francesco Sannini}{Conceptualization, Data curation, Investigation, Methodology, Software, Writing -- original draft, Validation, Visualization, Writing -- review \& editing}
\credit{ Francesco Salzano}{Data curation, Formal analysis, Methodology, Software, Visualization, Writing -- original draft, Writing -- review \& editing}
\credit{ Simone Scalabrino}{Methodology, Supervision, Writing -- review \& editing}
\credit{ Rocco Oliveto}{Supervision, Writing -- review \& editing}
\credit{ Remo Pareschi}{Supervision, Funding acquisition, Writing -- review \& editing}
\credit{ Corrado Aaron Visaggio}{Supervision, Writing -- review \& editing}
\credit{ Andrea Di Sorbo}{Conceptualization, Investigation, Methodology, Project administration, Supervision, Writing -- original draft, Writing -- review \& editing}

\definecolor{string}{RGB}{163,21,21}
\definecolor{keyword}{RGB}{0,0,255}
\definecolor{identifier}{RGB}{43,145,175}

\lstdefinelanguage{json}{
  basicstyle=\ttfamily\small,
  showstringspaces=false,
  breaklines=true,
  literate=
   *{:}{{{\color{keyword}:}}}{1}
    {,}{{{\color{keyword},}}}{1}
    {"role"}{{{\color{identifier}"role"}}}{1}
    {"content"}{{{\color{identifier}"content"}}}{1}
    {"messages"}{{{\color{identifier}"messages"}}}{1}
    {"}{{{\color{string}"}}}{1},
}

\tcbuselibrary{skins, breakable}
\def\BibTeX{{\rm B\kern-.05em{\sc i\kern-.025em b}\kern-.08em
    T\kern-.1667em\lower.7ex\hbox{E}\kern-.125emX}}

\journal{Journal of Systems and Software}

\begin{document}

\begin{frontmatter}



\title{Identifying and Characterizing Semantic Clones of Solidity Functions}


\author[inst1]{Ermanno Francesco Sannini\corref{cor1}}
\ead {e.sannini@studenti.unisannio.it}

\author[inst2]{Francesco Salzano}
\ead {francesco.salzano@unimol.it}

\author[inst2]{Simone Scalabrino}
\ead{simone.scalabrino@unimol.it}

\author[inst2]{Rocco Oliveto}
\ead{rocco.oliveto@unimol.it}

\author[inst2]{Remo Pareschi}
\ead{remo.pareschi@unimol.it}

\author[inst3]{Corrado Aaron Visaggio}
\ead{corrado.visaggio@unifg.it}

\author[inst1]{Andrea Di Sorbo}
\ead{disorbo@unisannio.it}

\cormark[*]
\cortext[cor1]{Corresponding author}

\affiliation[inst1]{
organization={Department of Engineering},
addressline={University of Sannio}, 
city={Benevento},
country={Italy}}

\affiliation[inst2]{
organization={Department of Biosciences and Territory},
addressline={University of Molise}, 
city={Pesche},
country={Italy}}

\affiliation[inst3]{
organization={Department of Agricultural Sciences, Food, Natural Resources and Engineering},
addressline={University of Foggia}, 
city={Foggia},
country={Italy}}

\begin{abstract}
Smart Contracts are essential blockchain components, mainly written in Solidity. The high availability of public Solidity code leads to frequent reuse and high clone ratios. 
Since cloning can propagate vulnerabilities and flaws, effective detection is crucial. Although existing techniques work well in detecting syntactic clones, the identification of semantic clones is an open problem. To address this challenge, in this paper, we present and empirically assess a scalable methodology, based on analyzing code and comments, to spot semantically equivalent Solidity functions. We first collected an up-to-date dataset of about 300,000 Ethereum smart contracts, 82.07\% of which are compliant with modern Solidity version 0.8. Manual validation of a statistically significant sample comprising 1,155 function pairs confirms the effectiveness of our solution, achieving an overall precision of 59\% (rising to 84\% for homonymous functions) and a recall of 97\%. Besides, we explore the structural differences occurring on semantically equivalent Solidity functions, demonstrating that they often represent design alternatives focused on security choices, modularization, and gas optimization. Finally, we investigate the use of Large Language Models (LLMs) as documentation engines in scenarios where code comments are poor or absent. Our results show that LLM-generated summaries, combined with sentence transformers like BERT, can bridge the documentation gap, enabling the identification of semantic clones in uncommented code with 75\% precision. This work establishes a modern benchmark for Solidity clone detection and provides a foundation for the automated discovery of secure and efficient code alternatives.
\end{abstract}



\begin{keyword}
Smart Contracts \sep  Solidity \sep Code Clones Detection \sep LLM


\end{keyword}

\end{frontmatter}



\section{Introduction}
\label{sec:intro}
Smart contracts (SCs) are executable programs aimed to automate, verify, and enforce the terms of an agreement between untrusted parties that run on top of a blockchain \cite{zou2019smart}. Once deployed, a SC is immutable, and its code is publicly accessible, meaning that everyone can read the code, understand it, and replicate it. Blockchain technology ensures the immutability of an SC, guaranteeing that transactions initiated by the contract are reliably executed. Several blockchain platforms support SCs \cite{wan2019programmers}, with Ethereum being the most prominent one \cite{alharby2017blockchain}. The significant attraction gained by the development of SCs in recent years is due to the potential of this technology to transform a wide range of industries, including financial services, supply chain management, and the IoT \cite{sun2023demystifying}. Solidity is the most popular SC high-level language \cite{hwang2020gap, wohrer2020domain}, supported by different blockchain platforms \cite{dhaiouir2020systematic}. 

Given the high availability of Solidity code, to avoid reinventing the wheel and reduce development efforts~\cite{wan2021smart}, developers tend to frequently reuse code~\cite{chen2021understanding}. Portions of code that are identical or similar to the original sources are known as \textit{code clones}~\cite{kondo2020code}. 
Depending on differences in code structure and behavior, clones are commonly classified into four types~\cite{roy2009comparison}: \textit{Type-1} are exact copies, \textit{Type-2} and \textit{Type-3} include partial modifications such as variable renaming, type changes, or statement edits, while \textit{Type-4} clones are semantically similar but implemented differently. 

High code clone ratios in Ethereum SCs have been found by previous research. For example, Kondo~et~al. observed that 79.2\% of the verified contracts are clones~\cite{kondo2020code}, while Khan~et~al. analyzed clones at the granularity of functions and reported a 30.1\% overall clone ratio, out of which 27.0\% are exact duplicates~\cite{khan2022code}.


However, this widespread practice of cloning in an immutable, high-stakes environment carries severe risks. As many SCs suffer from hidden flaws and vulnerabilities \cite{gao2019smartembed}, indiscriminate cloning can rapidly propagate security risks across the ecosystem \cite{he2020characterizing}. For a developer, the ability to identify a set of semantically equivalent implementations (Type-4 clones) is not merely an academic exercise; it is a crucial step toward making informed engineering choices. By comparing functionally identical alternatives, developers can proactively select implementations that offer better security, improved gas efficiency, or a more robust design, thereby mitigating critical smart contract flaws. Despite this need, the tools and resources available to support this task are severely lacking. While existing techniques can detect syntactic clones (i.e., Type-1, Type-2, and Type-3 clones), identifying semantic clones remains an open challenge \cite{zakeri2023systematic}. The state-of-the-art tool, EClone \cite{liu2018eclone}, not only suffers from low precision and high execution overhead but is also incompatible with SCs having an up-to-date Solidity version \cite{wang2025clone}. A critical gap in the literature compounds this problem: the lack of up-to-date, curated datasets for evaluating new techniques.



To address these limitations, this paper introduces and evaluates a lightweight and scalable method for identifying Type-4 clones of Solidity functions. The hypothesis behind our methodology is that developer-written code comments, which can foster a deeper understanding of function intent and behavior~\cite{rani2021identify}, can fit as a strong representative for their semantics.
Specifically, we suppose that functions with \textit{high} comment similarity and \textit{low} code similarity are good candidates as Type-4 clones.
To test this hypothesis, we conduct a manual validation on a statistically significant sample of function pairs extracted from a curated dataset of active and up-to-date SCs deployed on the Ethereum blockchain. 
We also characterize the structural and implementation-level differences of the validated clones, revealing common patterns that can inform better development practices.
Finally, since SC functions often lack code comments and Large Language Models (LLMs) have proven effective at automatically summarizing source code \cite{tian2023chatgpt}, we explore the potential of LLMs to generate comments useful for complementing the approach in cases of poorly documented functions.

The results of our study demonstrate that accounting for the (dis)similarity between both code and comments is viable for detecting semantically equivalent functions in the context of Solidity SCs. Furthermore, the approach successfully identifies functionally equivalent implementations for core behaviors like asset transfers and access control, enabling the characterization of the structural differences underlying these core behaviors. Finally, our results highlight the potential of using LLM-generated comments to address the common challenge of missing documentation.
While our results demonstrate promising effectiveness, this work does not provide a definitive or complete solution to the challenge of Type-4 clone detection in Solidity SCs. Instead, it should be interpreted as an empirical study aimed at better understanding the characteristics of semantically equivalent functions and the kinds of signals that can help their identification. Our findings highlight dimensions that could be leveraged by more structured and robust approaches in the future. Therefore, our contribution represents a foundation for pursuing the design of next-generation techniques, rather than a conclusive methodology for semantic clone detection in the context of SCs.

In summary, the main contributions of our paper are as follows:
\begin{itemize}
    \item \textit{we present an empirical study in which we test whether a joint analysis of source code and comments (both written by developers and automatically generated) can effectively help practitioners detect semantic clones of SCs;}
    \item \textit{we study how semantically-equivalent Solidity functions differ at both structural and implementation levels;}
    \item \textit{we provide a curated dataset featuring active and up-to-date SCs that can be used for research and practical purposes.} 
\end{itemize}


\subsection{Motivating Example}
\label{sec:motivation}

The identification of \textit{Type-4} clones is particularly challenging in the context of Solidity due to the flexibility of the language and the critical nature of SC security. Type-4 clones, or semantic clones, are code fragments that perform the same computation but are implemented using different syntactic structures. To illustrate the importance of detecting such clones, consider the two functions presented in Figure~\ref{fig:motivating_example}.

\begin{figure*}[htbp]
\centering
\begin{minipage}{0.48\textwidth}
\begin{lstlisting}[style=solidity, caption={Implementation A (Safe)}, label={lst:safe_transfer}]
function transfer(address _to, uint256 _amount) public { 
    require(balances[msg.sender] >= _amount); 
    balances[msg.sender] -= _amount; 
    (bool success, ) = payable(_to).call{value: _amount}(""); 
    require(success); 
}
\end{lstlisting}
\end{minipage}
\hfill
\begin{minipage}{0.48\textwidth}
\begin{lstlisting}[style=solidity, caption={Implementation B (Vulnerable)}, label={lst:vulnerable_transfer}]
function transfer(address _to, uint256 _amount) public { 
    Account storage userAccount = accounts[msg.sender]; 
    bool isEligible = userAccount.balance >= _amount ? true : false; 
    if (isEligible) { 
        uint256 previousBalance = userAccount.balance; 
        userAccount.balance = previousBalance - _amount; 
        payable(_to).call{value: _amount}(""); 
        assert(userAccount.balance == previousBalance - _amount); 
    } 
} 
\end{lstlisting}
\end{minipage}
\caption{Two semantically equivalent functions with different syntactic structures and security properties}
\label{fig:motivating_example}
\end{figure*}

Both functions in Figure~\ref{fig:motivating_example} share the same high-level semantic goal: deducting a specified amount from the sender's balance and transferring it to a target address. However, they differ in several key aspects:

\begin{itemize}
    \item \textbf{Syntactic Structure:} Listing~\ref{lst:safe_transfer} uses a simple mapping and a direct \lstinline|require| check. In contrast, Listing~\ref{lst:vulnerable_transfer} retrieves a storage \texttt{struct}, uses a ternary operator for eligibility, and wraps the logic in a conditional \texttt{if} block. Standard syntax-based clone detectors (Types 1-3) would likely fail to link these two functions due to the high degree of variation in tokens and AST structure.
    
    \item \textbf{Security and Correctness:} While Listing~\ref{lst:safe_transfer} correctly follows the \textit{Checks-Effects-Interactions} pattern and validates the return value of the low-level call, Listing~\ref{lst:vulnerable_transfer} contains a critical flaw. It fails to check the \texttt{bool success} return value of the \texttt{.call()} method. If the transfer fails (e.g., if the recipient contract rejects the payment), the function in Listing~\ref{lst:vulnerable_transfer} will still finalize the balance deduction, resulting in a loss of funds for the user.
\end{itemize}

This example highlights why semantic clone detection is crucial. If a vulnerability is discovered in one implementation, searching for identical or near-identical code is not enough. Developers must be able to identify a set of functions that perform the same semantic task, regardless of how they are syntactically implemented, to ensure that buggy logic is not replicated across different contracts. Our approach addresses this by leveraging natural language documentation and code embeddings to bridge the gap between intent and implementation.

\textbf{Paper structure.} The remainder of the paper is organized as follows. Section~\ref{sec:related} explores the related literature, highlighting the novelty of the proposed approach. Section~\ref{sec:design} describes the research questions, the collected dataset, and the methodology followed. Results are presented and discussed in Section~\ref{sec:results}, while their practical implications are reported in Section~\ref{sec:discussion}. Section~\ref{sec:threats} deals with threats to the study's validity, and, finally, Section~\ref{sec:conclusions} concludes the paper outlining future research directions.

\section{Related work}
\label{sec:related}
This section overviews the research contributions related to our work.
\subsection{Smart Contract Datasets}

Numerous SC datasets have been released to support empirical research on security and analysis tools. 
These resources differ substantially in size, annotation quality, and the range of vulnerabilities they cover. 
Durieux et al. conducted one of the earliest large-scale evaluations using the SmartBugs framework~\cite{durieux2020empirical,ferreira2020smartbugs}, analyzing tens of thousands of SCs and revealing high false-positive and false-negative rates across vulnerability detectors. 
Subsequent efforts aimed to provide more comprehensive datasets: Ibba et al. curated around 50,000 Solidity SCs enriched with structural metrics and vulnerability annotations from Slither~\cite{ibba2024curated}, while Zheng et al. released the DAPPSCAN datasets (source and bytecode) and benchmarked seven analysis tools using labels derived from audit reports~\cite{zheng2024dappscan}. 
Despite their value, most of these datasets suffer from two notable limitations. 
First, they typically overlook the presence of SC clones, which can bias empirical evaluations and inflate performance metrics. 
Second, the majority of SCs are implemented in outdated Solidity versions. 
A recent survey by Iuliano et al. covering 37 SC datasets found that most code samples rely on versions 0.4.x and 0.5.x~\cite{iuliano2025solidity}, limiting their applicability to modern development practices and tools.
Overall, while existing SC datasets have enabled significant progress in empirical research, their outdated nature and lack of clone-awareness reduce their suitability for contemporary SC analysis.

\subsection{Smart Contract Code Summarization and Comment Generation}

Recent advances in SC code summarization have combined syntactic and semantic representations to improve the generation of natural language descriptions of source code. 
Shi et al. proposed CoSS, a framework that leverages Control Flow Graphs and Graph Neural Networks to capture statement-level semantics~\cite{shi2023coss}. 
By jointly modeling control dependencies and token-level features, CoSS achieved superior performance over nine prior models across \textit{BLEU}, \textit{ROUGE-L}, \textit{METEOR}, and \textit{CIDEr} metrics on \textit{Java}, \textit{Python}, and \textit{Solidity} datasets.
Following such advances in SC summarization, Mao et al. introduced \textit{SCLA}, which enhances SC code summarization through LLMs and semantic enrichment techniques~\cite{mao2024scla}. 
\textit{SCLA} constructs Abstract Syntax Trees to extract latent semantic features such as callback structures and global variables, and generates enriched prompts for LLMs using semantically similar examples retrieved with \textit{SentenceTransformer}. 
Experiments on 14,795 function-comment pairs show that \textit{SCLA} significantly outperforms existing state-of-the-art models.
Similarly, Zhang et al. presented \textit{CCGRA}, a retrieval-enhanced comment generation approach that integrates a retrieval module with a pretrained \textit{CodeT5} model~\cite{zhang2023ccgra}. 
By constructing a cleaned dataset of 29,720 code-comment pairs and removing duplicates and template-based examples, \textit{CCGRA} achieved higher scores across \textit{BLEU}, \textit{ROUGE-L}, and \textit{METEOR} metrics, as well as improved human-rated comment naturalness and informativeness.

\subsection{Clone Detection}
\label{related:clonedetection}
The public availability of SC code on the blockchain facilitates reuse but also leads to the replication of insecure or vulnerable patterns~\cite{zou2019smart}. Consequently, detecting and characterizing SC clones has become an important research topic. 
Khan et al. analyzed 33,073 SCs to investigate cloning practices on Ethereum~\cite{khan2022code}, replicating and extending prior studies on Solidity-based projects. 
Earlier, Kondo~et~al. reported an unusually high clone rate (79.2\%)~\cite{kondo2020code}. 
A subsequent reanalysis at the function level (considering \textit{Type-1}, \textit{Type-2}, and \textit{Type-3} clones) revised this estimate to 30.13\%, of which 27.03\% were exact duplicates. 
Mo~et~al. further confirmed the widespread presence of cloning in SC ecosystems by analyzing 26,294 SCs comprising 97,877 functions~\cite{mo2025code}. 
Their study revealed that approximately one-third of clones (32.01\%) co-evolve and that, unlike traditional software, cloned SCs are rarely involved in bug-fixing tasks.
Pinku et al. compared ASTNN, GMN, and CodeBERT for semantic clone detection~\cite{pinku2024use}, underscoring the challenges of identifying semantic clones across programming languages such as Java, Python, C, and C++ using \textit{SemanticCodeBench}.
Similarly, CodeBERT has been evaluated for clone detection on several Java datasets~\cite{arshad2022codebert}. It performs best on \textit{Type-1} and \textit{Type-4} clones but generalizes poorly to unseen code, with recall dropping by up to 40\%. Fine-tuning mitigates this issue, improving recall by around 30\%. However, CodeBERT is not tailored for Solidity.
Mo et al. analyzed 26,294 SCs containing 97,877 functions, confirming the widespread presence of code clones in SCs~\cite{mo2025code}. Their study offered valuable insights into cloning behavior, notably revealing that 32.01\% of clones co-evolve. Unlike traditional software, SC clones are rarely involved in bug-fixing tasks.
While these studies demonstrate the progress achieved in semantic clone detection using deep learning and pretrained models, they primarily focus on general-purpose programming languages. 
In contrast, research specifically targeting SCs remains limited. 
In this field, Wang~et~al. evaluated five clone detection approaches for Solidity SCs~\cite{wang2025clone}. Among these, EClone~\cite{liu2018eclone} was the only technique able to detect Type-4 clones. Although their study was the first to address Type-4 clones in SCs and included a manual analysis of 380 contracts, the dataset was limited to contracts verified in July 2018~\cite{wang2025clone}.
EClone achieved a high recall (88.2\%) for Type-4 clones by analyzing EVM bytecode, but suffered from low precision (29.6\%) and limited scalability. While EClone is the only approach capable of dealing with \textit{Type-4} clones, it relies on \textit{Oyente}~\cite{luu2016making}, which is deprecated, and supports only Solidity versions up to 0.4.x. 
This limitation prevented us from employing EClone as a baseline in our study.

\section{Study Design}
\label{sec:design}
\begin{figure}[htb]
    \centering
    \includegraphics[width=\linewidth]{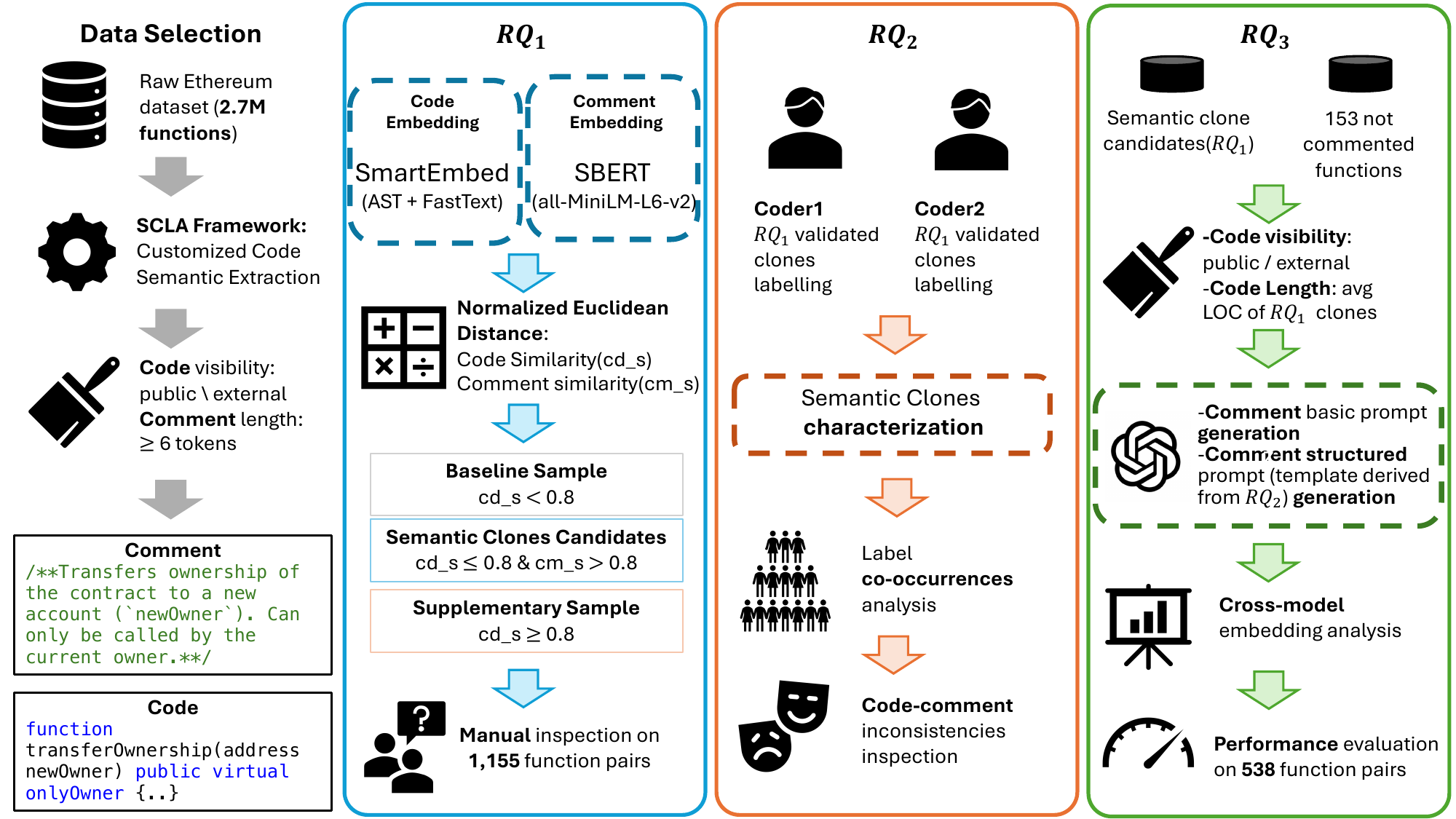} 
    \caption{Overview of the proposed methodology}
    \label{fig:workflow}
\end{figure}

In this section, we introduce the research questions that guided our work, the experimental dataset, and the methodology followed to answer them.

\subsection{Research Questions}

The \textit{goal} of our study is twofold: (i) to evaluate the effectiveness of combining code and comment analysis in identifying semantic clones of Solidity functions and (ii) to gain a better understanding of the prevalence and nature of structural and semantic variations within these clones across active Ethereum smart contracts. Based on this goal, we pose the following research questions:

\begin{itemize}
    \item \textbf{RQ$_1$}: \textit{How effective is an approach that analyzes both comments and code in identifying semantic clones of Solidity functions across active and up-to-date Ethereum smart contracts?} 
    \item \textbf{RQ$_2$}: \textit{What are the main structural and semantic differences occurring in Type-4 clones of Solidity functions?}
    \item \textbf{RQ$_3$}: \textit{To what extent can an LLM bridge the gap left by missing or incomplete comments in the identification of semantic clones?} 
\end{itemize}

RQ$_1$ drives our empirical study of semantic clones from a quantitative point of view, evaluating a method to identify them from commented Solidity functions. In RQ$_2$, we deeply inspect identified semantic clone pairs to depict their prominent traits, providing a qualitative perspective. Finally, in RQ$_3$, given that a considerable proportion of Solidity functions lack comments, we examine the impact of the function comments automatically generated by an LLM in the proposed semantic clones identification approach, supplying a strategy for poorly documented functions, too.

\textbf{Replication package.} 
All scripts, data, and results of the study are made publicly available in our replication package~\cite{dataset}

\subsection{Data Selection}
\label{sec:dataset}
The data acquisition process followed a multi-stage pipeline designed to ensure the relevance, uniqueness, and semantic richness of the analyzed SCs. This process, summarized in the first phase of the workflow overview in Figure~\ref{fig:workflow}, is detailed below.

\subsubsection{Contract Retrieval}
\begin{lstlisting}[
    language=SQL, 
    caption={Query used to collect smart contracts}, 
    label={lst:query}, 
    basicstyle=\scriptsize\ttfamily, % ttfamily rende il font monospaziato, tipico del codice
    keywordstyle=\color{blue}\bfseries, % mette in blu e grassetto le parole chiave SQL
    commentstyle=\color{green!50!black},
    stringstyle=\color{orange},
    breaklines=true, % va a capo automaticamente se la riga e troppo lunga
    showstringspaces=false
]
SELECT contracts.address, COUNT(*) AS tx_count, MAX(transactions.block_timestamp) AS last_transaction_time
  FROM `crypto_ethereum.contracts` AS contracts
  JOIN `crypto_ethereum.transactions` AS transactions ON (transactions.to_address = contracts.address)
  GROUP BY contracts.address
  ORDER BY tx_count DESC
\end{lstlisting}

Data collection followed the double-step methodology used by Durieux~et~al.~\cite{durieux2020empirical}, reproduced in the following.
First, we used \textit{Google BigQuery}~\cite{medvedev2018ethereum} on the public Ethereum dataset \textit{bigquery-public-data.crypto\_ethereum}\footnote{\url{https://cloud.google.com/blog/products/data-analytics/data-for-11-more-blockchains-in-bigquery-public-datasets}}, to retrieve all stored contract addresses and the number of their associated transactions, with the date of the last recorded transaction.
We adopted the query reported in Listing~\ref{lst:query}, which yielded $8,922,530$ contract.
To retrieve active and employed SCs only, all addresses related to contracts with fewer than $10$ transactions or whose last transaction was earlier than 2024 were filtered out, obtaining a smaller set of $272,014$ items.
Second, we used \textit{Etherscan}, a renowned platform to investigate the Ethereum blockchain~\cite{grech2018madmax}, to retrieve the source codes associated with the addresses retrieved in the first step.
However, not all contracts have their source code available on Etherscan, and among those that are accessible, many are duplicates. To address this, we applied a cleaning step on the $203,033$ files retrieved. After removing whitespaces and tabulations, we filtered out the duplicates having the same MD5 checksums. This filtering process reduced the dataset to $82,337$ files, whose statistics are summarized in Table~\ref{tab:dataset_stats}.

\begin{table*}[htb]
\centering
\caption{Dataset statistics}
\label{tab:dataset_stats}
\begin{tabularx}{\textwidth}{@{} X r @{}}
\toprule
\textbf{Metric} & \textbf{Value} \\
\midrule
Number of files                            & 82,337    \\
Number of contracts                        & 296,343   \\
Number of functions                        & 2,705,194 \\
Number of public and external functions    & 1,798,160 \\
Median function length                     & 138.0     \\
Longest function length                    & 56,047    \\
Shortest function length                   & 15        \\
Number of commented functions              & 660,895   \\
Median comment length                      & 105.0     \\
Longest comment length                     & 3,380     \\
Shortest comment length                    & 6         \\ 
\bottomrule
\end{tabularx}
\end{table*}


Length metrics (i.e., median, longest, and shortest function/comment length) refer to the number of characters, but the comment-related statistics are computed only on at least two-word comments, to drop the out-of-relevance ones. Notably, the percentage of uncommented functions is $75.57\%$. Regarding the Solidity versions used in the SCs featured in the dataset (see Table~\ref{tab:solidity-versions}), version $0.8$ is by far the most prevalent, adopted by $82.07\%$ of the contracts. This widespread use reflects the significant safety and functionality improvements introduced in that version, aligning with our decision to focus on up-to-date contracts. Older versions such as $0.4$, $0.5$, and $0.6$ still account for about $15\%$ of the dataset, often due to the complexity of upgrading legacy code. Version $0.7$, by contrast, is adopted by a minority of SCs ($1.80\%$ of the dataset), as it is generally easier to migrate from it to newer versions. SCs with no version are often meant to be used as libraries or dependencies; therefore, the compiler version of these utterly depends on the contract that uses them. 

\begin{table*}[htb] 
\centering
\caption{Distribution of Solidity versions}
\label{tab:solidity-versions}
\begin{tabularx}{\textwidth}{@{} X r r @{}}
\toprule
\textbf{Solidity version} & \textbf{Count} & \textbf{Percentage} \\
\midrule
0.4        & 19,779  & 6.67\%  \\
0.5        & 14,012  & 4.73\%  \\
0.6        & 11,611  & 3.92\%  \\
0.7        & 5,348   & 1.80\%  \\
0.8        & 243,262 & 82.07\% \\
No version & 2,331   & 0.79\%  \\
\bottomrule
\end{tabularx}
\end{table*}

\subsubsection{Function Extraction and Semantic Filtering}
The extraction of individual functions was performed using the SCLA framework, namely, an automated SC summarization tool via LLMs and control flow prompt \cite{mao2024scla}. We specifically leveraged and customized the \textit{Code Semantic Extraction Stage} to efficiently fetch relevant data without the overhead of full Abstract Syntax Tree (AST) generation. For every extracted function, we stored the following attributes in a structured CSV format: \textit{contract\_id, contract\_name, solidity\_version, contract\_variables, function\_name, function\_visibility, token\_length, function\_code}. In this context, \textit{function\_code} represents the function body purged of formatting characters, extra spaces, and inline comments, while \textit{function\_comment} specifically refers to the documentation located in the function header. From the initial set of $82,337$ unique files, we extracted a total of $2,705,194$ functions. To prepare the data for semantic clone detection, we applied three primary filtering criteria:
\begin{itemize}
\item \textbf{Visibility Constraints:} We focused exclusively on \textit{public} and \textit{external} functions, as these represent the primary interface of the SC and are the entry points for blockchain transactions, making them the most critical components for security and logic analysis.
\item \textbf{Documentation Presence:} As our methodology relies on natural language analysis for semantic identification, we removed the $75.57\%$ of functions that lacked header comments.
\item \textbf{Semantic Quality:} A preliminary manual inspection revealed that many comments were either syntactically trivial (e.g., \lstinline|'.', '+++'|) or lacked descriptive depth (e.g., \lstinline|'See @{IERC20-allowance}'|). To ensure the next step embeddings captured meaningful semantic intent, we discarded functions with comments shorter than six tokens.
\end{itemize}
After applying these filters, the final experimental dataset consisted of $216,429$ high-quality function pairs, providing a robust foundation for identifying Type-IV clones.

\subsection{Identification of Semantic Clones}
To answer $RQ_1$, we define a strategy for eliciting the functions candidate as semantic clones from the whole dataset. This process (see the blue box in Figure~\ref{fig:workflow}) involves defining equivalence criteria, generating embeddings, and executing a stratified manual validation.

\subsubsection{Criteria for Semantic Equivalence}

\begin{figure*}[htbp]
\centering
\begin{minipage}{0.48\textwidth}
\begin{lstlisting}[style=solidity]
function decreaseAllowance(address spender, uint256 subtractedValue) public returns (bool) 
{   _approve(msg.sender, spender, _allowances[msg.sender][spender].sub(subtractedValue)); 
    return true; }
\end{lstlisting}
\end{minipage}
\hfill
\begin{minipage}{0.48\textwidth}
\begin{lstlisting}[style=solidity]
function decreaseAllowance(address spender, uint subtractedValue) public virtual returns (bool)
{   address owner = _msgSender(); 
    uint currentAllowance = allowance(owner, spender); 
    require(currentAllowance >= subtractedValue, "ERC20: decreased allowance below zero"); 
    unchecked { _approve(owner, spender, currentAllowance - subtractedValue);
    } return true; }
\end{lstlisting}
\end{minipage}

\caption{Comparison of \textit{decreaseAllowance} function with one semantically equivalent}
\label{fig:type4-decreaseAllowance}
\end{figure*}

\begin{figure*}[htbp]
\centering
\begin{minipage}{0.48\textwidth}
\begin{lstlisting}[style=solidity]
function transferOwnership(address newOwner) public onlyOwner 
{   require(newOwner != address(0), "Ownable: new owner is the zero address");
    emit OwnershipTransferred(_owner, newOwner);
    _owner = newOwner;}
\end{lstlisting}
\end{minipage}
\hfill
\begin{minipage}{0.48\textwidth}
\begin{lstlisting}[style=solidity]
function transferOwnership(address newOwner) public onlyOwner {
    require(newOwner != address(0), "Ownable: new owner is the zero address, use renounceOwnership Function");
    emit OwnershipTransferred(owner, newOwner);
    if(balanceOf(owner) > 0) _basicTransfer(owner, newOwner, balanceOf(owner));
    setExemptions(owner, false, false, false, false);
    setExemptions(newOwner, true, true, true, false);
    owner = newOwner;}
\end{lstlisting}
\end{minipage}

\caption{Comparison of \textit{transferOwnership} function with one not semantically equivalent}
\label{fig:notype4-transferOwnership}
\end{figure*}

We labeled a pair of functions as a semantic clone only if they satisfy a strict set of equivalence criteria. The criteria include:
\begin{itemize}
\item \textbf{Signature Compatibility}: Functions must share identical input parameter counts, types, and return types.
\item \textbf{State-Change Equivalence}: Successful execution must result in identical modifications to the contract's core state variables (see Figure~\ref{fig:type4-decreaseAllowance}). We explicitly excluded pairs where one function implemented specialized logic, such as custom fee computations. For example, one \textit{transfer} function might execute a simple peer-to-peer value exchange, while another implements a fee-on-transfer mechanism. 
\item \textbf{Syntactic Tolerance}: We accepted variations that do not alter the semantic outcome, including the presence of auxiliary safety checks (e.g., \texttt{require} or \texttt{assert}), the use of functionally equivalent libraries (e.g., \textit{SafeMath} vs. native Solidity 0.8+ arithmetic), and the reordering of non-dependent instructions.
\end{itemize}
By adhering to these rules, our manual validation process was able to systematically and consistently distinguish true semantic clones from functions that were simply similar or related (see Figure~\ref{fig:notype4-transferOwnership}).

\subsubsection{Embedding Generation and Similarity Calculation}
We represent each function by generating vector representations (embeddings) for both the function code and its comments. In this way, we capture both the implementation logic and the developer's documented intent. 
We used \textit{SmartEmbed}~\cite{gao2020checking} to generate vector representations of the source code. This tool employs a custom parser built on the ANTLR4 Solidity grammar\footnote{ \url{https://github.com/solidityj/solidity-antlr4}} to analyze the syntactic structure. It analyzes the code syntactic structure (through an AST), extracts key elements (e.g., loops and conditions), and applies the FastText algorithm~\cite{bojanowski2017enriching}  after a text normalization step. 
For the comments, we employed \textit{SBERT}~\cite{reimers2020making}, specifically, the \textit{all-MiniLM-L6-v2}\footnote{ \url{https://huggingface.co/sentence-transformers/all-MiniLM-L6-v2}} model. This transformer-based model is optimized for semantic search and sentence similarity. Given that our dataset's mean comment length is $22.25$ tokens (median: 18), the model's 256-word window is more than sufficient to capture the full context of function headers without truncation.

Finally, we calculated the similarity between every two code fragments from their embeddings using the normalized Euclidean distance ~\cite{mikolov2013efficient}.
We established a classification strategy based on three distinct sets:
\begin{enumerate}
\item \textbf{Baseline Set} ($88.68B$ pairs): To evaluate the impact of comments and avoid selection bias, we first established a \textit{baseline} by analyzing the dataset across the code similarity range $[0, 0.8)$. This allowed us to assess the raw existence of semantic clones through a purely code-based search, providing the essential contrast for our proposed methodology. 
\item \textbf{Semantic Clone Candidates} ($2.74M$ pairs): Defined by \textit{low code similarity} ($\le 0.8$) and \textit{high comment similarity} ($> 0.8$), to isolate functions that are semantically identical (based on documentation) but syntactically distinct (based on code). This choice balances the literature threshold of $0.7$ ~\cite{wang2025clone} as the lower bound for syntactic clone detection with an adequate code distinction between pairs already below $0.8$ similarity, empirically retrieved. On the other hand, BERT-based cosine similarity approaches correlate well with human perception in code summarization tasks, supporting the use of similarity thresholds set to $0.7$–$0.8$ for detecting semantically similar comments~\cite{zhang2019bertscore}.
\item \textbf{Supplementary Set} ($8.2B$ pairs): Pairs with code similarity $\ge 0.8$, utilized to identify potential False Negatives, i.e. the cases where high syntactic similarity might mask Type-4 clones or lean toward Type-3.
\end{enumerate}
As literature suggests CodeBERT as a strong candidate for semantic clone identification (see Section \ref{related:clonedetection}), we conducted an additional baseline experiment using it on Semantic Clone Candidates and Supplementary Sample. However, the pre-trained model paid a minimum similarity of $0.9$, a value too high to effectively discriminate the structural variance typical of Type-4 clones in Solidity.

\subsubsection{Manual Validation and Statistical Soundness}
To ensure the statistical soundness of this multi-faceted evaluation, we conducted a manual validation on a statistically significant sample of $1,155$ pairs. This total was derived from three stratified samples of 385 items each, providing a $95\%$ confidence level and a $5\%$ margin of error \cite{maxwell2008sample}. Rather than a straightforward random selection, we employed a stratified sampling strategy to ensure that each sample accurately reflected the underlying distribution of the dataset. For the Semantic Clone Candidates, the sample was stratified based on both comment similarity (range $[0.8, 1]$) and code similarity (range $[0, 0.8]$), as detailed in Table~\ref{tab:type4-validation}. For the parts hiding the full code similarity spectrum, we defined specific similarity stripes: the supplementary sample (code similarity $\ge 0.8$) was divided into four stripes with a step of $0.05$ (e.g., $(0.95, 1]$, $(0.9, 0.95]$, etc.), while the baseline sample (code similarity $< 0.8$) was divided into four stripes with a step of $0.2$ (e.g., $[0, 0.2)$, $[0.2, 0.4)$, etc.). In all three samples, the number of items randomly chosen for each stripe was proportional to the percentage of pairs that each stripe represents of the total population.

The manual analysis was performed by two domain-expert coders, both authors of the paper. Both evaluators are Ph.D students with at least 5 years of experience in the IT industry.  To ensure consistency, starting from the Type-4 clone candidates sample, they conducted a pilot on 30 candidates randomly sampled from different stripes to define a common identification strategy and then independently inspected the remaining pairs.
Cohen’s Kappa~\cite{cohen1960coefficient} was used to measure the inter-rater agreement, achieving a score of 0.83, which indicates \textit{almost perfect agreement} before a conflict resolution phase on 30 couples. 
All conflicts were subsequently resolved through discussion; notably, 13 of these were finally classified as non-clones after the evaluators agreed to hold a more conservative view on specific functionalities (i.e., not considering as semantic clones instances in which specific behaviors were implemented, such as burning and fees operations).
The three-part validation process, using both the primary (low code similarity and high comment similarity) and baseline and supplementary samples (code similarity only), allowed us to derive True Positives (TP), False Positives (FP), False Negatives (FN), and True Negatives (TN).
Specifically, the manual validation of our primary sample yielded the count for TPs (i.e., the number of pairs confirmed as true clones) and FP (i.e., the number of pairs rejected as non-clones). Concurrently, the analysis of the other samples allowed us to quantify the FNs (i.e., true clones initially filtered out by our approach) and the TNs (i.e., non-clones that were correctly rejected).
With this complete confusion matrix, we can provide a robust evaluation of Precision, Recall (TP rate), F1-score, Specificity (TN rate), and Accuracy.

\subsection{Characterizing Semantic Clones}
To address $RQ_2$, we performed a qualitative characterization of the semantic clones validated during the $RQ_1$ phase (see the red box in Figure~\ref{fig:workflow}). The goal was to identify the recurring implementation patterns that distinguish Type-4 clones in Solidity.

\subsubsection{Labeling Procedure}
The labeling process was conducted by the same two domain experts who performed the $RQ_1$ validation. Following the conflict resolution phase, the experts analyzed the implementation differences between confirmed clone pairs and assigned one or more labels to each pair. This inductive approach led to the definition of some recurring patterns arising in a semantic modification of a function. Therefore, they explained these common formats into seven categories:
\begin{itemize}
\item \textbf{\textit{modifier}}: The functions differ in their use of Solidity modifiers (e.g., access control or custom state guards) within their signatures.
\item \textbf{\textit{safemath}}: One function utilizes the \textit{SafeMath} library for arithmetic overflow protection, while the other uses native operators (common in Solidity 0.8+).
\item \textbf{\textit{call\_super} / \textit{call\_internal}}: The functions differ in modularization, where one delegates logic to a superclass or an internal helper function, while the other implements it inline.
\item \textbf{\textit{diff\_algo}}: The functions employ fundamentally different algorithmic approaches or control flow structures to achieve the same semantic outcome.
\item \textbf{\textit{spec\_impl}}: Represents distinctive implementation choices or specific coding styles that are unique to the developer's context but semantically neutral.
\item \textbf{\textit{add\_check}}: One function includes additional validation logic (e.g., \texttt{require} or \texttt{assert} statements) that the other lacks, without altering the core state transition.
\end{itemize}
The experts used a shared, multi-column sheet to assign one or more labels to each validated pair, allowing for a multidimensional characterization of complex clones. For instance, a single pair might be labeled as both \textit{safemath} and \textit{add\_check} if it utilizes library-based arithmetic and includes additional input validation.

\subsubsection{Structural and Semantic Analysis}
To provide a more comprehensive view of the nature of semantic clones, we strengthened the manual labeling with three additional analyses. We first computed the percentage of validated clones that share the same function name versus those with distinct names. This analysis provides insight into how often semantic clones are hidden under different identifiers, which is a meaningful challenge for traditional search-based tools. Then, we assessed the prevalence of code-comment misalignment, where the natural language documentation does not accurately reflect the underlying code logic. This metric measures the reliability of documentation-based detection and quantifies the noise in our methodology. Eventually, during the manual inspection, we performed a qualitative review of the function headers. This allows us to characterize the general state of documentation in the Solidity ecosystem, further contextualizing the SBERT results.
By combining these qualitative labels with quantitative metrics on naming and alignment, we provide a comprehensive overview of the diversity and prevalence of Type-4 clones in active Ethereum SCs.

\subsection{Bringing the Documentation Gap with LLM}
A significant challenge identified in Section~\ref{sec:dataset} is that $75.57\%$ of Solidity functions lack the documentation required for our primary detection pipeline. $RQ_3$ investigates whether Large Language Models (LLMs) can effectively bridge this gap by generating semantic summaries that serve as proxies for developer-written comments (see the green box in Figure~\ref{fig:workflow}).

\subsubsection{Preliminary Evaluation: Direct LLM Classification}
We first investigated whether an LLM could directly act as a Type-4 clone detector. We conducted a zero-shot experiment on the $RQ_1$ validation sample, using \textit{ChatGPT-4o}. The model was prompted (see Listing~\ref{lst:prompt_4_clones}) to classify pairs as semantic clones with a binary "YES/NO" response, with the temperature set to 0 to ensure reproducibility.

\begin{lstlisting}[language=json, label=lst:prompt_4_clones, caption=Prompt used to detect semantic clones., basicstyle=\scriptsize\ttfamily]
{"role": "system", "content": "You are a helpful software engineering assistant, your task is to say if the Solidity functions are semantic clones."},
{"role": "user", "content": f"Here are the two code functions:
First function: {first_artifact}
Second function: {second_artifact}
Your response is just YES or NO."}
\end{lstlisting}

The model's performance was notably poor, identifying only 43 out of 385 pairs (11\%) as semantic clones. Furthermore, a manual inspection revealed that 14 of these 43 identified pairs were Type-3 clones, suggesting the model's struggle to distinguish high syntactic similarity and true semantic equivalence. These findings justified our decision to shift toward a more robust strategy: using the LLM for code summarization rather than direct classification.

\subsubsection{LLM-based Summarization Strategy}
To generate the missing semantic context, we leveraged \textit{ChatGPT-4o} using the prompt: \texttt{Create a summary(no additional feedback) of this Solidity function: <function-source-code>}. The choice of such a model was endowed by the study of Wang~et~al.~\cite{wang2023zero}, which exhibited ChatGPT as the state-of-the-art in zero-shot cross-lingual summarization, without the need for fine-tuning. They also found that ChatGPT generally produces verbose summaries with straightforward information. This, together with the common misleading and mispleading nature of comments, steered our choice for a very basic prompt, with the \textit{no additional feedback} specification to hold verbosity and stress model flexibility. Moreover, we checked the LLM-generated comments while manually checking the sample, ensuring the correctness of these generated comments. 
To evaluate the effectiveness of these generated comments, we compared three distinct transformer-based embedding models: \textit{all-MiniLM-L6-v2}, already used on the real comments of $RQ_1$; \textit{BERT base}\footnote{ \url{https://huggingface.co/google-bert/bert-base-uncased}},  trained by Google on a large corpus of English data for either Masked Language Modeling or Next Sentence Prediction~\cite{DBLP:journals/corr/abs-1810-04805}; \textit{CodeBERT base}\footnote{ \url{https://huggingface.co/microsoft/codebert-base}}, even trained on source code by Microsoft for Masked Language Modeling as well as Replaced Token Detection~\cite{feng2020codebert}. These values were used in combination with code similarity values computed in $RQ_1$ to investigate whether comments generated by the LLM can be used instead of real comments, also hinting at how three well-known pre-trained models perform. 

\subsubsection{Targeted Sampling}
To assess the applicability of this approach to real-world scenarios, we selected a subset of the 100 most active Ethereum contracts (by transaction volume) from the dataset introduced in Section~\ref{sec:dataset}. From these, we extracted all public and external functions. As an additional filter, we decided to take care of function length. Referring to the average number of terms associated with the clones validated in $RQ_1$, we selected only the functions that were at least 26 words long. For this subset, we focused on hidden clones: functions with the same name (homonymous) but low code similarity ($cd\_s < 0.8$). We generated summaries for these pairs with the same procedure introduced above, and then we computed the comment similarity using SBERT. Finally, we manually validated all pairs with a comment similarity $> 0.8$ to confirm their Type-4 clone status.

\subsubsection{Prompt Refinement with Structured Summarization}
\label{sec:rq2:sub:prompt_refinement}
Finally, we attempt to improve the quality of LLM-generated documentation by incorporating the findings from $RQ_2$. We developed a structured prompt that required the LLM to follow a specific plain-text template: \texttt{Response plain-text template: a brief overview of the function’s purpose, implementation details, and noteworthy behaviors, brief parameters description, a brief description of the return value.} This structured approach aims to force the model to capture the facets of the function (e.g., specific implementation choices and modifiers) identified during our manual characterization phase, leading to more accurate semantic embeddings and higher detection accuracy for uncommented functions.

\section{Results}
\label{sec:results}
In this section, we document the results of our study by
responding to the three research questions that we posed.

\begin{table*}[htb!]
\centering
\scriptsize
\caption{Summary of manual validation results across Baseline, Semantic clone candidates, and Supplementary samples}
\label{tab:type4-validation}
\resizebox{\textwidth}{!}{
\begin{tabular}{
  p{3.5cm}
  p{3cm}
  r
  r
  r
  r
  r
  r
}
\toprule
\textbf{Comment Similarity (cm\_s)} & \textbf{Code Similarity (cd\_s)} & \textbf{Stripe Size} & \textbf{\% for Stripe} & \textbf{Candidates} & \textbf{Val. Rate} & \textbf{Same Name} & \textbf{Val. Rate$'$} \\
\midrule
\multicolumn{8}{c}{Semantic clone candidates} \\
\midrule
\multirow{4}{*}{$0.95< \text{cm\_s} \leq 1.00$} 
& $0.60< \text{cd\_s} \leq 0.80$ & 755,738 & 27.55\% & 106 & 71\% & 95 & 75\% \\
& $0.40< \text{cd\_s} \leq 0.60$ & 193,771 & 7.06\% & 27 & 63\% & 23 & 70\% \\
& $0.20< \text{cd\_s} \leq 0.40$ & 67,977 & 2.48\% & 10 & 40\% & 4 & 100\% \\
& $0.00< \text{cd\_s} \leq 0.20$ & 830 & 0.03\% & 0 & - & 0 & - \\
\midrule
\multirow{4}{*}{$0.90< \text{cm\_s} \leq 0.95$}
& $0.60< \text{cd\_s} \leq 0.80$ & 138,403 & 5.05\% & 19 & 74\% & 16 & 88\% \\
& $0.40< \text{cd\_s} \leq 0.60$ & 26,600 & 0.97\% & 4 & 50\% & 4 & 50\% \\
& $0.20< \text{cd\_s} \leq 0.40$ & 3,697 & 0.13\% & 1 & 100\% & 1 & 100\% \\
& $0.00< \text{cd\_s} \leq 0.20$ & 23 & 0.00\% & 0 & - & 0 & - \\
\midrule
\multirow{4}{*}{$0.85< \text{cm\_s} \leq 0.90$}
& $0.60< \text{cd\_s} \leq 0.80$ & 1,000,083 & 36.46\% & 140 & 56\% & 80 & 95\% \\
& $0.40< \text{cd\_s} \leq 0.60$ & 61,071 & 2.23\% & 9 & 78\% & 7 & 71\% \\
& $0.20< \text{cd\_s} \leq 0.40$ & 20,692 & 0.75\% & 3 & 67\% & 2 & 100\% \\
& $0.00< \text{cd\_s} \leq 0.20$ & 198 & 0.01\% & 0 & - & 0 & - \\
\midrule
\multirow{4}{*}{$0.80< \text{cm\_s} \leq 0.85$}
& $0.60< \text{cd\_s} \leq 0.80$ & 417,367 & 15.22\% & 58 & 50\% & 31 & 94\% \\
& $0.40< \text{cd\_s} \leq 0.60$ & 37,226 & 1.36\% & 5 & 60\% & 3 & 100\% \\
& $0.20< \text{cd\_s} \leq 0.40$ & 19,186 & 0.70\% & 3 & 0\% & 0 & - \\
& $0.00< \text{cd\_s} \leq 0.20$ & 185 & 0.01\% & 0 & - & 0 & - \\
\midrule
\textbf{Total} & – & \textbf{2,743,047} & – & \textbf{385} & \textbf{59\%} & \textbf{266} & \textbf{84\%} \\
\midrule
\multicolumn{8}{c}{Baseline sample} \\
\midrule
\multirow{4}{*}{-}
& $0.60< \text{cd\_s} \leq 0.80$ & 44,573,449,757 & 50.26\% & 194 & 1\% & - & - \\
& $0.40< \text{cd\_s} \leq 0.60$ & 19,215,400,058 & 21.67\% & 83 & 0\% & - & - \\
& $0.20< \text{cd\_s} \leq 0.40$ & 5,087,296,833 & 5.74\% & 22 & 0\% & - & - \\
& $0.00< \text{cd\_s} \leq 0.20$ & 19,805,198,292 & 22.33\% & 86 & 0\% & - & - \\
\midrule
\textbf{Total} & – & \textbf{88,681,344,940} & – & \textbf{385} & \textbf{$<$ 1\%} & \textbf{-} & \textbf{-} \\
\midrule
\multicolumn{8}{c}{Supplementary sample} \\
\midrule
\multirow{4}{*}{-}
& $0.95< \text{cd\_s} \leq 1.00$ & 1,048,024,727 & 12.78\% & 49 & 2\% & - & - \\
& $0.90< \text{cd\_s} \leq 0.95$ & 1,297,854,132 & 15.83\% & 61 & 3\% & - & - \\
& $0.85< \text{cd\_s} \leq 0.90$ & 2,450,097,921 & 29.88\% & 115 & 3\% & - & - \\
& $0.80< \text{cd\_s} \leq 0.85$ & 3,404,977,654 & 41.52\% & 160 & 0\% & - & - \\
\midrule
\textbf{Total} & – & \textbf{8,200,954,434} & – & \textbf{385} & \textbf{2\%} & \textbf{-} & \textbf{-} \\
\bottomrule
\end{tabular}}
\end{table*}

\subsection{\texorpdfstring{\textbf{RQ$_1$}}{RQ1}: Effectiveness of the proposed approach}

To assess the effectiveness of the approach based on the joint analysis of code and comments to identify semantic code clones of Solidity functions in active and up-to-date Ethereum SCs,  we conducted a manual validation study on $1,155$ function pairs, divided into three distinct representative samples of $385$ items each. As we can see in Table~\ref{tab:type4-validation}, the Type-4 Candidates sample was selected across a grid of comment-code similarity ranges. Each row in this section corresponds to a specific stripe, defined by a bin of comment similarity (range $[0.8, 1.0]$) and code similarity (range $[0.0, 0.8]$). Conversely, the Baseline and Supplementary samples were stratified on code similarity only to provide a comparison group and to identify false negatives. 
The \textit{Comment Similarity} column indicates the ranges used to define each sampling stripe, based on the similarity between natural language comments (e.g., function descriptions), as the column \textit{Code Similarity} does with the corresponding code implementations. \textit{Stripe Size} reports the number of clone candidates in the dataset that fall within a specific combination of comment similarity and code similarity values, as defined by the corresponding similarity range. The value of \textit{\% for Stripe} shows the proportion that this stripe represents relative to the entire dataset. \textit{Candidates} denotes the number of pairs of clones manually assessed within each stripe, and \textit{Val. Rate} reflects the percentage of confirmed as true \textit{Type 4} clones through manual validation. The \textit{Same Name} column counts how many validated clones had identical function names, often suggesting shared intent or functionality. 
The column \textit{Val.\ Rate$'$} captures how many candidates were confirmed as clones and also had equivalent function names. The last two concerns are valid only for Type 4 clones.

\subsubsection{Stripe Analysis}
As expected, the largest stripe sizes appear in the highest ranges of comment similarity, particularly for $0.80 < \textit{cm\_s} \leq 0.90$, $0.95 < \textit{cm\_s} \leq 1.00$, and $0.80 < \textit{cm\_s} \leq 0.85$ with code similarity $\textit{cd\_s}$ between $0.60$ and $0.80$, accounting for 36.46\%, 27.55\%, and 15.22\% of the total, respectively. For these ranges, validation rates are in that order 56\%, 71\%, and 50\%, which grew to 95\%, 75\%, and 94\%, respectively, considering only homonymous functions. However, stripes with $\textit{cd\_s} \leq 0.40$ show very few or no candidates. Overall, on the sample of 385 candidates manually validated, the approach analyzing both comments and code achieved a Precision of 59\%. If we consider the 266 pairs with the same name, the Precision grows up to 84\%.
In severe contrast, the manual validation of the baseline sample (pair selection based on low code similarity only, ignoring comments) identified a single Type-4 clone, resulting in a validation rate of approximately 0.5\%. This negligible value is a straightforward measure of the impact that comments have on semantic clones identification.

\subsubsection{Qualitative Analysis}
On the other hand, the samples not confirmed during manual validation convey several discriminating factors.  Only five pairs of those validated as actual Type-4 clones have a different name ($2\%$ of the validated ones). 
Having a high comment similarity for these pairs is justified by an incorrect comment (e.g., \textit{allowance} with the same comment of \textit{transferOwnership}) or a comment that differs only by one term (i.e., the comments of the functions \textit{increaseAllowance} and \textit{decreaseAllowance} vary only by the terms increases and decreases). Other non-clone pairs were 18 Type-3 clones (5\% of the sample), which differ either by a different \textit{modifier} or by the use of the \textit{safemath} library instead of basic mathematical operators. Notable is the case of three pairs in which a known function was used for a different task (e.g., \textit{decreaseAllowance} used to change the contract owner, and \textit{increaseAllowance} to exploit safe and block lists). Eventually, there was a set of pairs with an evident difference in behavior: distinct management of tax computation; redirections of tokens to hardcoded wallet addresses; and unconventional control flow using custom flags and collections. All the given examples can be examined in the replication package.

\subsubsection{Global Performance Metrics}
To achieve a comprehensive performance evaluation, we manually validated a third representative sample, which was designed to identify false negatives (FN). The analysis of these 385 pairs (see Table~\ref{tab:type4-validation}, Supplementary sample), which our approach initially excluded due to high code similarity, revealed only seven FN. The validation process was exhaustive, involving the resolution of six conflicts among the raters, which were attributed to borderline cases or human error and settled after discussion.
With the FN count established, alongside the TP and FP derived from our primary sample, we computed a Recall of $97\%$, demonstrating the approach's effectiveness in capturing nearly all existing semantic clones, and a F1-Score of $74\%$. Furthermore, we obtained a Specificity of $71\%$, showing a good capability to correctly identify non-clones, and an overall Accuracy of $79\%$.

\subsubsection{Clones Distribution}
\begin{table*}[htb]
\centering
\caption{Top 20 Type-4 cloned functions ordered by number of occurrences}
\label{tab:top20_functions_clones}
\begin{tabularx}{\textwidth}{@{} X r @{}}
\toprule
\textbf{Functions} & \textbf{Candidates} \\
\midrule
transferFrom        & 29,276,384 \\
transferOwnership   & 13,115,324 \\
increaseAllowance   & 11,420,559 \\
transfer            & 3,665,680  \\
decreaseAllowance   & 1,881,371  \\
renounceOwnership   & 531,743    \\
burnFrom            & 487,766    \\
approve             & 377,605    \\
owner               & 233,361    \\
setApprovalForAll   & 163,248    \\
revokeRole          & 130,214 \\
grantRole           & 126,871 \\
name                & 94,366 \\
symbol              & 90,706 \\
burn                & 68,643 \\
mint                & 63,553 \\
safeTransferFrom    & 63,124 \\
totalSupply         & 50,717 \\
balanceOf           & 42,539 \\
pause	           & 31,216 \\
\bottomrule
\end{tabularx}
\end{table*}


As further analysis, Table~\ref{tab:top20_functions_clones} shows the first 20 functions for candidate semantic clones number, applying the proposed approach to homonymous pairs in the whole dataset. The top entries \textit{transferFrom} (29M), \textit{transferOwnership} (13M), and \textit{increaseAllowance} (11M) reveal a huge frequency of cloning in functions handling token and permission management. These are core ERC-20 and ERC-721 functions, so it is not stunning their duplicated use across many projects, making them the best candidates for fine adaptations and semantic changes.
Functions like \textit{decreaseAllowance}, \textit{renounceOwnership}, and \textit{burnFrom}, basic in handling tokens’ supply, ownership, and delegation, exhibit hundreds of thousands of clones, too. This corroborates the common practices in SC design, where even little variations in allowance handling or ownership transfer require code adaptation~\cite{he2020characterizing}.
Specialized functions such as \textit{safeTransferFrom}, \textit{revokeRole}, and \textit{grantRole} also appear among the most cloned. Thus, role-based access control and safe token transfer mechanisms, although not primary transfer functions, still demonstrate considerable usage in practice. This emphasizes the need for robust access control and secure item transactions as critical components of SC development~\cite{ghaleb2023achecker}.
Finally, Type-4 clones often target core contract operations, such as asset transfers, permission handling, and ownership management, using a wide range of distinct approaches that ultimately achieve the same functionality.

\smallskip
\noindent
\fcolorbox{black}{lightgray}{
\begin{minipage}[center]{0.98\linewidth}
\textbf{RQ$_1$ Summary}: The joint analysis of code and natural language documentation proves highly effective in identifying semantic clones. Out of the 385 candidates manually assessed in our primary sample, 59\% were confirmed as true Type-4 clones; this precision significantly increases to 84\% when focusing on homonymous functions. The methodology demonstrates high robustness with a Recall of 97\%, an F1-score of 74\%, and an overall Accuracy of 79\%. Our findings confirm that semantic cloning is a widespread practice in the Ethereum ecosystem, particularly within core functionalities such as asset transfer (\textit{transferFrom}), ownership management (\textit{transferOwnership}), and access control (\textit{increaseAllowance}), where standardized intent is frequently implemented through specific code structures.
\end{minipage}}
\smallskip

\subsection{\texorpdfstring{\textbf{RQ$_2$}}{RQ2}: Type-4 code clones characteristics}

\begin{table*}[htb]
\centering
\caption{Occurrences of the top three labels assigned for single value, in pairs, and in triplets}
\label{tab:top_labels}
\begin{tabularx}{\textwidth}{@{} X X r r @{}}
\toprule
\textbf{Group} & \textbf{Label} & \textbf{Count} & \textbf{Percentage} \\
\midrule
\multirow{5}{*}{Single value} 
  & call\_internal & 50 & 22\% \\
  & diff\_algo     & 37 & 16\% \\
  & add\_check     & 22 & 10\% \\
  & spec\_impl     & 7  & 3\% \\
  & call\_super    & 6  & 3\% \\
\midrule
\multirow{13}{*}{Pairs} 
  & modifier, call\_internal & 22 & 10\% \\
  & modifier, add\_check     & 20 & 9\% \\
  & safemath, call\_internal & 19 & 8\% \\
  & modifier, safemath       & 17 & 7\% \\
  & safemath, add\_check     & 12 & 5\% \\
  & modifier, diff\_algo     & 8  & 4\% \\
  & call\_internal, diff\_algo & 8 & 4\% \\
  & modifier, call\_super       & 7 & 3\% \\
  & safemath, diff\_algo        & 7 & 3\% \\
  & call\_internal, spec\_impl  & 4 & 2\% \\
  & call\_internal, add\_check  & 3 & 1\% \\
  & diff\_algo, add\_check      & 3 & 1\% \\
  & call\_super, add\_check     & 2 & 1\% \\
\midrule
\multirow{5}{*}{Triplets} 
  & modifier, call\_internal, spec\_impl & 4 & 2\% \\
  & modifier, safemath, call\_internal   & 4 & 2\% \\
  & safemath, call\_internal, diff\_algo & 3 & 1\% \\
  & modifier, safemath, add\_check       & 2 & 1\% \\
  & modifier, call\_super, add\_check    & 2 & 1\% \\
\bottomrule
\end{tabularx}%
\end{table*}

This section explores the structural and semantic facets of the validated Type-4 clones, identifying the underlying development patterns that lead to functionally equivalent but syntactically diverse implementations.

The distribution of labels assigned to validated clones is summarized in Table~\ref{tab:top_labels}. For each label, Table~\ref{tab:top_labels} provides the count of Type-4 clones and the respective percentage of distribution in the case of a single or multiple (pair or triplet) assignment. Indeed, each clone can be associated with multiple labels, revealing different facets of code modifications.

The most frequent label is \textit{call\_internal}, appearing in 22\% of the cases, which suggests that many Type-4 clones involve the decomposition of complex logic into multiple, smaller internal method calls. Such refactoring operations typically aim to improve code readability, modularity, or maintainability by isolating specific functionality into reusable components, rather than duplicating entire code blocks.
The label \textit{diff\_algo} (16\%) is particularly noteworthy, as it indicates that the two functions achieve the same goal using different algorithms. Pairs tagged with such a label exhibit a \textit{mean code similarity} of $0.695$ and \textit{mean comment similarity} of $0.894$, showing that although these clones have notable structural differences,  they share high documentation-level similarities. The following most common label is \textit{add\_check} (10\%), indicating distinctions in validation logic by adding checks and \textit{require} statements. Less frequent are \textit{spec\_impl} (3\%) and \textit{call\_super} (3\%), which point to clones differentiated by specific implementations and superclass function invocations, respectively.

\subsubsection{Co-occurrence Analysis}
To achieve deeper insight, we also examined the co-occurrences of label pairs and triplets. Among the label pairs, the most common are \textit{modifier, call\_internal} (10\%), \textit{modifier, add\_check} (9\%), and \textit{safemath, call\_internal} (8\%). These combinations suggest that, in many clones, developers not only isolate logic into internal functions but also combine this refactoring with additional checks (\textit{add\_check}) or changes to access control (\textit{modifier}). Such a finding unveils the SC development effort to modularize and enhance the code at the same time.

The analysis of triplet co-occurrences further stressed this idea. For instance, triplets like \textit{modifier, call\_internal, spec\_impl} (2\%) and \textit{modifier, safemath, call\_internal} (2\%) reveal that developers combine multiple strategies within the same clone instance. These triplets often depict extensive modification not just for security or maintainability reasons but also for slight improvements, such as better suitability with coding standards (\textit{spec\_impl}) or fine access control enforcement.

Eventually, this co-occurrence analysis emphasizes that Type-4 clones are seldom the result of simple changes. Still, they often affect substantial semantic modifications that blend best practices, security hardening, and specific fitting. Overall, these findings highlight that semantic clones of Solidity functions often preserve the prevalent program behavior while substantially improving maintainability, safety, and flexibility.

\subsubsection{Representative Case Studies}

\begin{figure*}[htbp]
\centering
\begin{minipage}{0.48\textwidth}
\begin{lstlisting}[style=solidity]
/**
 * @dev See {IERC20-transferFrom}.
 * Emits an {Approval} event indicating the updated allowance. This is not required by the EIP. 
 * See the note at the beginning of {ERC20}.
 * NOTE: Does not update the allowance if the current allowance
 * is the maximum `uint256`.
 * Requirements:
 * - `from` and `to` cannot be the zero address.
 * - `from` must have a balance of at least `amount`.
 * - the caller must have allowance for ``from``'s tokens of at least `amount`.
 */
function transferFrom(address from, address to, uint256 amount) public virtual override returns (bool) 
{    address spender = _msgSender();
    _spendAllowance(from, spender, amount);
    _transfer(from, to, amount);
    return true;}
\end{lstlisting}
\end{minipage}
\hfill
\begin{minipage}{0.48\textwidth}
\begin{lstlisting}[style=solidity]
/**
 * @dev See {IERC20-transferFrom}.
 * Emits an {Approval} event indicating the updated allowance. This is not required by the EIP. 
 * See the note at the beginning of {ERC20}.
 * Requirements:
 * - `sender` and `recipient` cannot be the zero address.
 * - `sender` must have a balance of at least `amount`.
 * - the caller must have allowance for `sender`'s tokens of at least `amount`.
 */
function transferFrom(address sender, address recipient, uint256 amount) public virtual override returns (bool)
{   _transfer(sender, recipient, amount);
    uint256 currentAllowance = _allowances[sender][_msgSender()];
    require( currentAllowance >= amount,
        "ERC20: transfer amount exceeds allowance");
    _approve(sender, _msgSender(), currentAllowance - amount);
    return true;}
\end{lstlisting}
\end{minipage}
\caption{Comparison of \textit{transferFrom} function with one semantically equivalent}
\label{fig:type4-transferFrom}
\end{figure*}

An illustrative instance of a \textit{Type 4} clone is represented by two semantically equivalent implementations of the \texttt{transferFrom} function. As shown in Figure~\ref{fig:type4-transferFrom}, both versions enforce the same logic by transferring tokens from one account to another while respecting the allowance mechanism. The first implementation retrieves the caller with \texttt{\_msgSender()} and separates the allowance deduction using \texttt{\_spendAllowance} before executing the transfer through \texttt{\_transfer}. The second implementation performs the transfer first and then verifies and updates the allowance using direct access to \texttt{\_allowances} and a call to \texttt{\_approve}. Although the order of operations and the internal function calls differ, both implementations ensure that a transfer occurs only if the spender has sufficient allowance, making the two versions functionally equivalent despite structural differences. Such a pair of Type-4 clones comes with two different comments with a semantic similarity of $0.85$, indicating a good alignment.

\begin{figure*}[htbp]
\centering
\begin{minipage}{0.48\textwidth}
\begin{lstlisting}[style=solidity]
/**
 * @dev Returns the address of the current owner.
 */
function owner() public view virtual returns (address o) 
{   bytes32 slot = _OWNER_SLOT;
    assembly { o := sload(slot) }
\end{lstlisting}
\end{minipage}
\hfill
\begin{minipage}{0.48\textwidth}
\begin{lstlisting}[style=solidity]
/**
 * @dev Returns the address of the current owner.
 */
function owner() public view virtual returns (address) 
{   return _owner;}
\end{lstlisting}
\end{minipage}
\caption{Comparison of \textit{owner} function with one semantically equivalent}
\label{fig:type4-owner}
\end{figure*}

To add depth, we also present a clone pair that does not refer to the \textit{ERC-20} standard. Hence, as depicted in Figure~\ref{fig:type4-owner}, both versions return the same semantic value, thus the current owner of the contract, by following two syntactically different approaches. The version on the left employs inline assembly to load the address of the owner from a storage slot (\textit{\_OWNER\_SLOT}), which could be useful in proxy and upgradable contracts to avoid storage collisions, granting compatibility among different versions. On the other hand, the second variant directly accesses a state variable (\textit{\_owner}), following a more conventional implementation. In this case, the two clones present completely equivalent comments.

\subsubsection{Comment Coherence and Completeness}
As part of our analysis, we examined how frequently function comments failed to reflect actual behavior, which serves as an indicator of the robustness of our approach. Since the manual analysis also involved comment inspection, this allowed us to assess the overall coherence of comments in Solidity functions. Hence, from our sample, we observed that 15\% of comments were not coherent (i.e., if code and comments are connected to each other \cite{rani2023decade}), while other comments were not complete(i.e., the comment content is sufficiently detailed to support development and maintenance tasks). The comment incoherence is presumably due to copy-and-paste errors. 

\begin{lstlisting}[style=solidity, caption={Example of transfer function with an unstructured incoherent comment.}, label={lst:incoherent}]
/**
 * @dev Returns the symbol of the token, usually a shorter version of the name.
 */
function transfer(address recipient, uint256 amount) 
    public virtual override returns (bool) 
{   _transfer(_msgSender(), recipient, amount);
    return true;}
\end{lstlisting}

\begin{lstlisting}[style=solidity, caption={Example of transferOwnership function with an unstructured coherent comment.}, label={lst:incomplete}]
/**
 * @dev Transfers ownership of the contract to a new account (`newOwner`).
 * Can only be called by the current owner.
 */
function transferOwnership(address newOwner) public onlyOwner 
{   require(newOwner != address(0), "Ownable: new owner is the zero address, use renounceOwnership Function");
    emit OwnershipTransferred(owner, newOwner);
    if(balanceOf(owner) > 0) _basicTransfer(owner, newOwner, balanceOf(owner));
    setExemptions(owner, false, false, false, false);
    setExemptions(newOwner, true, true, true, false);
    owner = newOwner;}
\end{lstlisting}

For instance, Listing~\ref{lst:incoherent} reports a \texttt{transfer} function with an incoherent comment, while Listing~\ref{lst:incomplete} depicts an instance with a coherent, but not complete, comment. Indeed, the comment catches the main purpose of the function, however, it fails to provide details, as it does not explain that balances are transferred to \texttt{newOwner} and does not mention the \texttt{setExemptions} function. 

\smallskip
\noindent
\fcolorbox{black}{lightgray}{
\begin{minipage}[center]{0.98\linewidth}
\textbf{RQ$_2$ Summary}: Type-4 clones in Solidity are characterized by functional decomposition, algorithmic variance, and enhanced validation, often within ERC-20/721-specific adaptations that employ multiple refactoring strategies simultaneously. While these modifications improve modularity and security, the 15\% misalignment rate between comments and code highlights significant documentation challenges and the prevalence of copy-paste debt in the ecosystem.
\end{minipage}}
\smallskip

\subsection{\texorpdfstring{\textbf{RQ$_3$}}{RQ3}: Using an LLM to identify Type-4 code clones}

\begin{table*}[htb]
\centering
\caption{Performance of semantic clone (true positive -TN) no-clone (true negative-TN) detection for LLM generated comments of functions with similarity higher than 0.8 for distinct pre-trained models}
\label{tab:llm_sim}
\begin{tabularx}{\textwidth}{@{} X X X X X X @{}}
\hline
\multicolumn{2}{l}{\textbf{\textit{all-MiniLM-L6-v2}}} & 
\multicolumn{2}{l}{\textbf{\textit{BERT base}}} & 
\multicolumn{2}{l}{\textbf{\textit{CodeBERT base}}} \\ 
\hline
\textbf{TP} & \textbf{TN} & 
\textbf{TP} & \textbf{TN} & 
\textbf{TP} & \textbf{TN} \\ 
\hline
31\% & 84\% & 85\% & 28\% & 93\% & 0\% \\ 
\hline
\end{tabularx}
\end{table*}

To evaluate if a general-purpose LLM can support semantic clone detection of functions with poor or absent comments, we first generated summaries for the functions of the sample used in $RQ_1$ employing ChatGPT-4o. We compared their embeddings generated through three pre-trained SBERT models, measuring comment similarity and evaluating whether a similarity higher than $0.8$ corresponded to semantic clones. As shown in Table~\ref{tab:llm_sim}, the choice of the embedding model significantly dictates the detection behavior:
\begin{itemize}
\item\textit{all-MiniLM-L6-v2} (Conservative): This model correctly identified 84\% of non-clones (True Negatives) but only 31\% of true clones. Its high specificity suggests it is best suited for scenarios where avoiding False Positives is critical.
\item\textit{BERT-base} (Balanced): BERT achieved a much higher sensitivity, identifying 85\% of true clones. While its True Negative rate was lower (28\%), it proved more capable of capturing the semantic variation in LLM-generated summaries.
\item\textit{CodeBERT-base} (Over-permissive): Despite its training on source code, CodeBERT identified 93\% of clones but failed to detect any non-clones (0\% True Negatives). The model overestimated semantic similarity in nearly all cases, making it ineffective for discrimination in this context.
\end{itemize}

\subsubsection{Detecting Hidden Clones in Uncommented Functions}
Afterward, we considered the functions without comments from the top 100 SCs by transactions in our dataset. After filtering the non-homonymous and syntactic clones (i.e., SmartEmbed similarity higher than $0.8$), we obtained 153 function pairs. Replicating the procedure used for the commented functions and, as the absence of comments, using the LLM-generated summarize as function documentation, only four pairs exceeded the similarity threshold using \textit{all-MiniLM-L6-v2}, whereas \textit{BERT base} pinpointed 109 such pairs, of which 82 (75\%) were confirmed as semantic clones. This result is meaningful, as these clones were previously invisible due to the lack of developer comments. However, for 13 pairs the similarity computed on the generated comments is lower than $0.8$, even if they are semantic clones (12\% of false negatives), confirming its limited utility for low-density LLM summaries. 

\subsubsection{Impact of Prompt Engineering and Documentation Quality}
As an additional investigation, we compared the standard Base Prompt (the one used in the previous section) against a more sophisticated Response Template (as illustrated in section \ref{sec:rq2:sub:prompt_refinement}) for the 153 uncommented function pairs. Our analysis revealed a phenomenon of similarity saturation. When using the structured template, \textit{BERT base} assigned similarity scores above $0.8$ to nearly the entire sample ($150$ out of $153$ pairs). This suggests that while structured prompting produces more formal documentation, it also introduces structural noise, leading to an overestimation of semantic similarity.

\begin{lstlisting}[style=solidity, caption={Source code of the \textit{approve} function related to Table \ref{tab:comments}}, label={lst:funcom}]
function approve(address _spender, uint256 _value) public returns (bool)
{   require((_value == 0) || allowed[msg.sender][_spender]== 0);
    allowed[msg.sender][_spender] = _value;
    emit Approval(msg.sender, _spender, _value);
    return true;}
\end{lstlisting}

\begin{table*}[htb]
\centering
\caption{Real vs ChatGPT generated comments for \textit{approve} function}
\label{tab:comments}
\scriptsize
\renewcommand{\arraystretch}{0} 
\begin{tabular}{
  p{6.1cm}   
  p{3.9cm} 
  p{6.4cm}   
}
\toprule
\textbf{Real comment} & \textbf{Base prompt} & \textbf{Prompt with response template} \\
\midrule

\begin{lstlisting}[breaklines=true, breakindent=0.5pt]
@dev Approve the passed address to spend  the specified amount of tokens on behalf of msg.sender. Beware that changing an allowance with this method brings the risk that someone may use both the old and the new allowance by unfortunate transaction ordering. One possible solution to mitigate this race condition is to first reduce the spender's allowance to 0 and set the desired value  afterwards: https://github.com/ethereum/EIPs/issues/20#issuecomment-263524729
@param _spender The address which will spend the funds.
@param _value The amount of tokens to be spent.
\end{lstlisting}
&

\begin{lstlisting}[breaklines=true, breakindent=0.5pt]
The approve function allows the caller to set the _value allowance for _spender only if the current allowance is zero or the new value is zero, emits an Approval event, and returns true on success.
\end{lstlisting}
&

\begin{lstlisting}[breaklines=true, breakindent=0.5pt]
The approve function allows an address (msg.sender) to approve another address (_spender) to spend a specified amount (_value) of tokens on its behalf. The function requires that the current allowance is either zero or the value being set is zero. It then updates the allowance and emits an Approval event to log the approval action.
_spender: The address that is allowed to spend tokens on behalf of the caller.
_value: The number of tokens the spender is allowed to transfer.
The function returns a boolean value true to indicate that the approval operation was successful.
\end{lstlisting}
\\
\bottomrule
\end{tabular}
\end{table*}

The qualitative differences between these documentation styles are illustrated in Table~\ref{tab:comments}:
\begin{itemize}
\item \textbf{Real Comments}: These represent the gold standard of developer-written documentation. They are characterized by specific details, including safety warnings (e.g., reentrancy or front-running risks) and specific parameter mappings that reflect the original developer's intent.
\item \textbf{Base Prompt Summaries}: These LLM-generated summaries focus strictly on the function's core logic. While less detailed than real comments,  they provide a straightforward fingerprint that conducts  similarity scoring without the overhead of structural templates.
\item \textbf{Structured Template Summaries}: These follow a rigid format (Overview, Implementation, Parameters, Return). While they appear more professional and complete, they emerges less effective for automated clone detection, as the boilerplate text obscures the semantic features of the function.
\end{itemize}


\smallskip
\noindent
\fcolorbox{black}{lightgray}{
\begin{minipage}[center]{0.98\linewidth}
\textbf{RQ$_3$ Summary}: LLM-generated comments support semantic clone detection in both commented (85\% true positives) and uncommented (75\%) functions. The choice of model for generating comment embeddings, if a conservative option such as all MiniLM L6 v2 or a more permissive one like BERT base, along with prompt engineering, has a significant impact on clone detection. While structured templates improve documentation formality, they often lead to similarity saturation, suggesting that a base prompt approach is more effective to uncover hidden Type-IV clones.
\end{minipage}}
\smallskip

\section{Discussion}
\label{sec:discussion}
Identifying semantic clones is an active challenge in traditional software engineering, but this problem has become even more critical in the domain of smart contracts. The public and immutable nature of blockchains encourages a culture where code reuse is common practice. 
However, when a contract is deployed, any inherited logic, including its flaws, is permanent. Our work addresses this by demonstrating that a lightweight, twofold approach can effectively bridge the semantic gap that syntax-based tools cannot absolve.

\subsection{A Catalog of Alternatives}
The findings from $RQ_2$ suggest that semantic clones in Solidity should not be considered merely as redundant code, but as meaningful design alternatives. When a developer looks for the implementation of a standard operation, such as \textit{transferFrom} or \textit{ownershipControl}, our methodology provides a way to discover various functionally equivalent implementations. For a practitioner, this is not just a search tool but a risk-mitigation strategy. By selecting a functional equivalent, developers can proactively avoid well-known pitfalls, such as inheriting security vulnerabilities, inefficient gas consumption patterns, or inadequate code structures. Our categorization effectively provides the rationale for a catalog of actionable design alternatives, allowing developers to select the implementation that best fits their specific requirements.


\subsection{The Role of LLMs in Documentation Pipeline}
A key contribution of this work ($RQ_3$) is the integration of LLMs to solve the documentation gap. A critical consideration of our methodology is the computational cost associated with LLM inference. With 75.57\% of our dataset (approximately 2.04 million out of 2,705,194 functions) lacking comments, a brute-force summarization of every function would require an excessive amount of computational effort. To ensure the approach remains lightweight and scalable, we clarify that LLM-based comment generation is not a global pre-processing task. Instead, it is a targeted enrichment step triggered only when specific criteria are met. In practice, the LLM is invoked only for function pairs that have already passed a set of preliminary cheap heuristics, such as similar function names, compatible signatures, and high structural dissimilarity (low code-similarity scores). By applying the LLM-based comment generation strategy only to these high-probability candidates, the processing volume is reduced from millions of functions to a controllable subset of potential semantic clones.

Furthermore, our investigation into prompt engineering reveals a critical trade-off. For automated detection, the Base Prompt is superior. By focusing strictly on core behavior, it avoids the similarity saturation caused by the boilerplate text of structured templates, providing a cleaner semantic fingerprint for embedding models. For code comprehension, the Structured Template Prompt acts as a valid tool for generating coherent, human-readable documentation.

\subsection{Implications and Benchmarking}
Our work provides significant contribution to both SC researchers and developers. First, we provide a large-scale, up-to-date dataset where 82.07\% of SCs are compliant with Solidity version 0.8, addressing a well-documented gap in the literature~\cite{iuliano2025solidity}. 
Second, our lightweight approach establishes a scalable baseline for a problem where the state-of-the-art tool EClone~\cite{liu2018eclone} is heavyweight and outdated~\cite{wang2025clone}, representing an initial benchmark for up-to-date Solidity. This benchmark establishes that neither purely syntactic searches nor models like CodeBERT have the necessary discriminative ability for Type-4 detection, as they either fail to capture semantic intent or exhibit significant similarity bias. By leveraging developer comments (both real and generated), we provide a foundation for future, more advanced models that must balance semantic depth with computational efficiency.

\section{Threats to validity}
\label{sec:threats}
\unskip
\textit{Threats to construct validity}. Construct validity threats stand out in the reliance on SmartEmbed to build the set of clone candidates from which we extracted the sample used for the manual analysis, as it was released in 2019, and Solidity has continued to evolve. However, this point is mitigated through the manual validation of a statistically relevant sample of candidate clones. The same approach mitigates all the other construct validity threats related to the models involved in our study. Moreover, to get the comment similarity, we used the cosine similarity between embeddings, which might produce erroneous outcomes. However, this cosine similarity is used by a large number of studies across several topics and is widely accepted \cite{liu2018eclone,haque2022semantic,hong2022commentfinder}.


\textit{Threats to internal validity}. A threat to our work is the choice of the code similarity threshold lower than $0.8$ to identify Type-4 clones. Clone analysis literature~\cite{wang2025clone} agrees to employ a similarity higher than $0.7$ to identify syntactic clones. The choice to raise the threshold arose from having empirically noted that pairs with a similarity lower than $0.8$ are already quite different. Previous findings~\cite{hu2021automating} revealed the conciseness of SC functions (with a mean of $6.34$ number of lines of code), impacting relevant similarity differences for small code distinctions. However, our approach could identify Type-3 clones in the $0.7-0.8$ range. Indeed, during the manual inspection of $RQ_1$, we identified 18 \textit{Type-3} clones ($5\%$ of the sample with a mean code similarity of $0.777$ and a mean comment similarity of $0.966$). We extracted the portion of the sample with a code similarity higher than $0.7$ (42 pairs) and tested the clone detection capability of an existing clone detection tool. Our choice fell on \textit{Solidity-NiCad}\footnote{ \url{https://github.com/eff-kay/solidity-nicad}}, the NiCad clone detector handled Solidity language (validated on $33,073$ SCs extracted from Etherscan.io \cite{khan2022code}), which, in the experiment made by Wang~et~al.~\cite{wang2025clone}, assessed a precision of $96.3$ and a recall of $97.3$ on Type-3 clones. Of the 42 function pairs, it was able to detect only one Type-3 clone. 
We also employed NiCad capability on the 109 uncommented function samples used in $RQ_3$, and no clones were detected, corroborating the soundness of the methodology proposed. 
The manual validation may be a potential threat, too. Nevertheless, the risk of error is reduced by using a double and independent analysis, with two experienced evaluators who reached a high inter-rater agreement represented by a Cohen's kappa value of $0.83$.

\textit{Threats to external validity}. The dataset used in this study might not reflect the actual scenario if considering all blockchains and programming languages for SC development. Indeed, our analysis is limited to Solidity SCs running on Ethereum. Although Solidity is the predominant language for defining SCs~\cite{wohrer2020domain}, and a large portion of SCs across different blockchain platforms are written in Solidity, further research is needed to investigate the generalizability of the approach to SCs running on different blockchain platforms and developed with other languages.

\vspace{-2mm}
\section{Conclusion}
\label{sec:conclusions}

In this paper, we introduced a novel methodology for identifying Type-4 (semantic) clones in Solidity smart contracts, addressing one of the most challenging problems in code clone detection. Our method is based on finding function pairs that have low code similarity but high semantic similarity in their comments. 
Our empirical evaluation, conducted on a statistically significant sample from a modern dataset of nearly 300,000 Ethereum SCs, demonstrated the effectiveness of the approach. Manual validation confirmed that the identified clones often represent critical design alternatives for core contract operations, such as asset transfers and permission management. These alternatives often reflect engineering trade-offs involving security concerns, modularization, and gas optimization. To address the significant documentation gap, where 75.57\% of functions lack developer comments, we successfully integrated a targeted LLM-based summarization pipeline. We demonstrated that while direct LLM-based detection is currently unfeasible, LLMs are highly effective as proxy documentation engines, enabling the identification of hidden semantic clones in uncommented functions.

The implications of this work are twofold: it provides developers with a methodology to discover design alternatives for critical smart contract logic, and it establishes a scalable, modern benchmark for the research community. As future work, we propose improving comment quality with more advanced LLM-based approaches, such as few-shot prompts or RAG and fine-tuning techniques, eventually exploiting the existing set of semantically equivalent commented functions. Besides, a deeper analysis of the semantic clones can help define best practice guidelines for security, gas optimization, and code readability.

\section*{Declaration of generative AI and AI-assisted technologies in the writing process}

During the preparation of this work, the authors used Grammarly and ChatGPT/Gemini to improve the English and support the writing process, respectively. After using these tools/services, the authors reviewed and edited the content as needed and took full responsibility for the content of the published article.

\section*{CRediT authorship contribution statement}
\insertcreditsstatement

\section*{Data Availability}

We publicly release our replication package~\cite{dataset}, in which we provide our datasets, the scripts for building, and everything needed to replicate all the results of our experiment.

\section*{Declaration of competing interest}
The authors declare no competing interests.

\section*{Acknowledgements}
This work is partially funded by PRIN Project \textit{Trust Machines for TrustlessNess (TruMaN): The Impact of Distributed Trust on the Configuration of Blockchain Ecosystems} (Identifier Code 2022F5CLN2– CUP H53D23002400006) financed by the Italian Ministry of University. 





\bibliographystyle{elsarticle-num} 
\bibliography{biblio}






\end{document}